\documentclass[twocolumn,longbibliography]{revtex4-1}%{revtex4-2}

\usepackage{subfig,braket,slashed,url,multirow,amsmath,amsfonts,xcolor}
\usepackage[T1]{fontenc} % if needed
\allowdisplaybreaks[1]

\definecolor{myblue}{rgb}{0.33,0.30,0.85}
\definecolor{myred}{rgb}{1.02,0,0.05}

\usepackage[
colorlinks=true, 
pdfstartview=FitV, 
linkcolor=myblue, 
citecolor=myred,%magenta, 
urlcolor=myblue, 
bookmarks=true,
bookmarksnumbered=true,
pdftitle={},
pdfauthor={}
]{hyperref}
\newcommand{\+}{\dagger}
\newcommand{\ra}{\rightarrow}

\newcommand{\up}{\uparrow} 
\newcommand{\down}{\downarrow}

\newcommand{\pp}{\perp}  
\newcommand{\pr}{\parallel} 
\newcommand{\hyp}{\mathchar`-}

\newcommand{\U}{\mathrm{U}}
\newcommand{\SU}{\mathrm{SU}}
\newcommand{\SO}{\mathrm{SO}}

\newcommand{\1}{{\mathbf 1}}

\newcommand{\cR}{{\mathcal R}}

\newcommand{\cN}{{\mathcal N}}
\newcommand{\cS}{{\mathcal S}}

\begin{document}

\title{
Hall viscosity in the A-phase of superfluid $^3$He
}

\author{Takuya Furusawa}
\email{furusawa@stat.phys.titech.ac.jp}
\author{Keisuke Fujii}
\author{Yusuke Nishida}
\affiliation{Department of Physics, Tokyo Institute of Technology, Ookayama, Meguro, Tokyo 152-8551, Japan}

\date{\today}

\begin{abstract}
We construct the effective field theory for the A-phase of superfluid $^3$He up to the next-to-leading order in the derivative expansion.
To this end, we gauge the internal global symmetries of the theory on the curved space by introducing the background gauge fields and spatial metric so as to expose a hidden local symmetry known as the nonrelativistic diffeomorphism.
The nonrelativistic diffeomorphism is particularly useful to yield an additional constraint on the effective field theory and reveal a universal expression for the Hall viscosity in the A-phase.
We find it five orders of magnitude larger than that in the B-phase under a magnetic field so that its experimental observation is more feasible by measuring the induced elliptic polarization of sound waves.
\end{abstract}

\maketitle

%\tableofcontents

%---------------------------------------------
\section{Introduction}
Recent developments in the topological material science originate from the discovery of the quantum Hall systems~\cite{vonKlitzing:1980pdk, Tsui:1982yy}.
The topological properties are encoded into their time-reversal odd dissipationless transport coefficient known as the Hall conductivity~\cite{Thouless:1982zz}.
The Hall viscosity~\cite{Avron:1988aug, Hoyos:2014pba} is a viscous analog of the Hall conductivity and possibly appears in three-dimensional anisotropic liquids with broken time-reversal symmetry as well as in time-reversal odd liquids in two dimensions.
It is related to an adiabatic response to deformations of the spatial geometry~\cite{Avron:1995fg, Read:2010epa}
and attracts considerable attention as an analogous index to the Hall conductivity that can distinguish topological phases.
Theoretical studies have been performed in various systems, such as integer~\cite{Avron:1995fg,Biswas:2013ads} and fractional quantum Hall systems~\cite{Levary:1997ads, Read:2008rn, Tokatly_2009, Haldane:2009ads, Hoyos:2011ez, Abanov:2013ads, Abanov:2014ula}, topological insulators~\cite{Kimura:2010ads, Hughes:2011hv, Hughes:2012vg, Hassan:2015ads}, chiral superfluids and superconductors~\cite{Hoyos:2013eha, Moroz:2014ska, Shitade:2014iya}, the superfluid $^3$He B-phase (Balian-Werthamer state)~\cite{Fujii:2016mbc}, and so on~\cite{Saremi:2011ab, Chen:2011fs,Barkeshli:2011ads, Wiegmann:2013hca, Wiegmann:2019ads, Souslov:2019ads, Hsiao:2020ads}.
Although a number of its observable signatures has been proposed~\cite{
    Lucas:2014sia,
    Sherafati:2016ads,
    Scaffidi:2017ads,
    Ganeshan:2017ads,
    Banerjee:2017ads,
    Tuegel:2017ads, 
    Delacretaz:2017yia,
    Holder:2019ads},
the Hall viscosity has rarely been measured in experiments except for colloidal chiral fluid~\cite{Soni:2018ads} and graphene~\cite{Berdyugin:2019ads}.

Toward further experimental measurements of the Hall viscosity, it is still meaningful to determine the Hall viscosity in other systems and predict how it appears in observable quantities.
Here, we turn to the superfluid $^3$He A-phase (Anderson-Brinkman-Morel state)~\cite{Vollhardt:2013goo}.
This phase is realized at zero temperature under a sufficiently large magnetic field, where $^3$He atoms form spin-triplet $p$-wave Cooper pairs spinning around a spontaneously fixed axis.
Thanks to their intrinsic angular momentum,
the ground state spontaneously breaks the time-reversal and space-rotation symmetries, so that the A-phase naturally satisfies the conditions for a nonvanishing Hall viscosity.
Therefore, one may expect a larger Hall viscosity in the A-phase compared to the B-phase, where the above conditions are satisfied only by applying an external magnetic field~\cite{Fujii:2016mbc}.

In spite of such an expectation, it is theoretically difficult to analyze the A-phase quantitatively. This is because $^3$He atoms are strongly interacting, so that the perturbative expansion is no longer valid.
Hence, a reliable approach beyond the perturbative expansion is highly desired.
We here employ the effective field theory based on the symmetries and the systematic derivative expansion following Refs.~\cite{Son:2005rv, Hoyos:2013eha, Fujii:2016mbc}.
This approach is valid at zero temperature and thus applicable to the A-phase under a sufficiently large magnetic field.
The effective field theory is constructed on low-energy degrees of freedom, such as Nambu-Goldstone bosons and gapless fermions, and is expanded systematically in terms of their derivatives.
Because its form is constrained by the symmetries of the system,
it is important to recognize all the available symmetries of the A-phase.

To this end, it is advantageous to construct the effective field theory in the presence of the background gauge fields and the background spatial metric.
The benefit is that such background fields expose a hidden local symmetry known as the nonrelativistic diffeomorphism~\cite{Son:2005rv}.
Importantly, the nonrelativistic diffeomorphism yields an additional constraint on the effective field theory that remains even when the background fields are turned off.
We will find that the resulting universal expression for the Hall viscosity in the A-phase is a natural three-dimensional generalization of that in the chiral superfluid in two dimensions~\cite{Hoyos:2013eha, Moroz:2014ska, Shitade:2014iya}.
Furthermore, it will turn out to be five orders of magnitude larger than that in the B-phase under a reasonable magnetic field~\cite{Fujii:2016mbc}.
We emphasize that our predictions are based solely on the symmetries and the systematic derivative expansion, so that they are nonperturbative, model-independent, and quantitatively reliable.

This paper is organized as follows.
In Sec.~\ref{sec: eft}, we start our discussion by clarifying the global symmetries, building blocks, and power counting in the A-phase. We then construct the effective Lagrangian up to the next-to-leading order in the derivative expansion.
Subsequently, we compute the particle number current and the stress tensor from the constructed effective Lagrangian and determine the universal expression for the Hall viscosity in the A-phase in Sec.~\ref{sec: hall}.
Furthermore, we estimate the elliptical polarization of sound waves induced by the Hall viscosity, which serves as an experimental observable.
Finally, Sec.~\ref{sec summary} is devoted to a summary of our results.
Throughout this paper, we work in real-time formalism.
Spacetime and spatial indices are denoted as $\mu,\nu = t,x^1,x^2,x^3$ and $i,j = x^1,x^2,x^3$, respectively.
Spin indices are given by $s,s'=\ \up,\down$, and the Pauli matrices acting on them are given by $\sigma^\alpha$ ($\alpha = 1,2,3$).
Summations over repeated indices are implicitly understood.

%---------------------------------------------
\section{Effective field theory} \label{sec: eft}
%---------------------------------------------
\subsection{Symmetries}
Microscopically, $^3$He atoms are described by nonrelativistic spin $1/2$ fermions $\psi_s(x)$ of mass $m$ with a density-density interaction
and enjoy the global continuous symmetries~\cite{Fujii:2016mbc}:
\begin{equation}
    \U(1)_\cN \times \SU(2)_\cS \times G_\mathrm{spacetime},
\end{equation}
where $\U(1)_\cN$ is associated with the particle number conservation and $\SU(2)_\cS$ is the spin-rotation symmetry.
The spacetime symmetry $G_\mathrm{spacetime}$ includes the $\SO(3)$ space rotation, spacetime translation, and Galilean invariance.
Note that the microscopic theory also enjoys the time-reversal and parity symmetries.
Throughout this paper, we neglect the dipole-dipole interaction between $^3$He atoms.

For later convenience, we couple the system to background fields and promote the global symmetries to local ones.
To this end, we introduce the $\U(1)_\cN$ and $\SU(2)_\cS$ gauge fields, $A_\mu(x)$ and $B_\mu(x) = H(x) \delta_{\mu, t}+\delta B_\mu(x)$, as well as the spatial metric $g_{ij}(x) = e^a_i(x) e^a_j(x)$ so as to place the system on a curved spatial manifold.
Here, $H(x)$ and $\delta B_\mu(x)$ represent an applied magnetic field and a perturbation around it, respectively, and the vierbein $e^a_i(x)$ defines a local orthogonal coordinate ($a,b = 1,2,3$).
$A_\mu(x)$ and $B_\mu(x)$ give us the $\U(1)_\cN$ and $\SU(2)_\cS$ gauge invariance:
\begin{subequations}
    \begin{align}
        \psi_s(x) &\ra e^{i \lambda(x)}U_{ss'}(x)\psi_{s'}(x),
        \\
        A_\mu(x) &\ra A_\mu(x) + \partial_\mu \lambda(x),
        \\
        B_\mu(x) &\ra U(x)B_\mu(x)U^\+(x) - i \partial_\mu U(x) U^\+(x),
    \end{align}
\end{subequations}
where $e^{i \lambda(x)} \in \U(1)_\cN$ and $U(x) \in \SU(2)_\cS$.
On the other hand, $e^a_i(x)$ yields the local rotational $\SO(3)_\cR$ invariance: $e^a_i(x) \ra R_{ab}(x) e^b_i(x)$ ($R(x) \in \SO(3)_\cR$).
Furthermore, thanks to the spacetime-dependent metric $g_{ij}(x)$,
the system enjoys the infinitesimal nonrelativistic diffeomorphism~\cite{Fujii:2016mbc}:
\begin{subequations}\label{eq: nr-diff}
    \begin{align}
        (t, x^i) &\ra (t, x^i + \xi^i (x)),
        \\
        \psi(x) &\ra \psi(x) - \xi^k(x) \partial_k \psi(x),
        \\
        A_t(x) &\ra A_t(x) - \xi^k(x) \partial_k A_t(x)- \partial_t \xi^k(x) A_k(x),
        \\ \nonumber
        A_i(x) &\ra A_i(x) - \xi^k(x) \partial_k A_i(x) - \partial_i \xi^k(x) A_k(x) 
            \\
            &\quad - m g_{ij}(x) \partial_t \xi^j(x),
        \\
        B_t(x) &\ra B_t(x) - \xi^k(x) \partial_k B_t(x)- \partial_t \xi^k(x) B_k(x),
        \\
        B_i(x) &\ra B_i(x) - \xi^k(x) \partial_k B_i(x) - \partial_i \xi^k(x) B_k(x),
        \\
        e^a_i(x) &\ra e^a_i(x) - \xi^k(x) \partial_k e^a_i(x) - \partial_i \xi^k(x) e^a_k(x),
        \\ \nonumber
        g_{ij}(x) &\ra g_{ij}(x) - \xi^k(x) \partial_k g_{ij}(x) 
        \\
        &\quad - \partial_i \xi^k(x) g_{kj}(x)- \partial_j \xi^k(x) g_{ik}(x).
    \end{align}
\end{subequations}
The nonrelativistic diffeomorphism invariance is larger than the spatial general coordinate invariance introduced by a time-independent metric and is regarded as a local extension of the Galilean invariance~\cite{Son:2005rv}.
Therefore, after gauging the global symmetries, $^3$He atoms enjoy the following local continuous symmetries:
\begin{equation}
    \U(1)_\cN \times \SU(2)_\cS \times \SO(3)_\cR \times G_\mathrm{NR\hyp diffeo}.
\end{equation}

We then explain low-energy degrees of freedom in the A-phase, which serve as ingredients for the effective field theory.
The ground state in the A-phase spontaneously breaks the internal symmetries as
\begin{equation}\label{eq: ssb}
        \U(1)_\cN \times \SO(3)_\cR \times \SU(2)_\cS 
        \ra \U(1)_{\cN - \cR} \times \U(1)_{\cS}.
\end{equation}
While the spin symmetry breaks down to its $\U(1)$ subgroup, the former two symmetries break down to the subgroup $\U(1)_{\cN - \cR}$, which is generated by simultaneous rotations of $\U(1)_\cN$ and $\SO(3)_\cR$ as clarified below.
Thus, the A-phase hosts five gapless Nambu-Goldstone bosons 
thanks to the spontaneous symmetry breaking.
Besides, Bogoliubov quasiparticles in the A-phase are gapless at two points in the momentum space, which are described by left- and right-handed Weyl fermions~\cite{Vollhardt:2013goo}.

\subsection{Building blocks}
Let us introduce the Nambu-Goldstone bosons and describe the building blocks of our effective field theory in the boson sector.
In the presence of fluctuations associated with the Nambu-Goldstone bosons, the gap function, or the superfluid order parameter, takes the form of
\begin{equation}\label{eq: coset rep}
    \braket{\psi_{s}(x) \partial_a \psi_{s'}(x)}
    \propto 
    e^{2 i \theta(x)} R_{ab}(x) \phi^b_{0} [(d_\alpha(x)\sigma^\alpha) i \sigma^2]_{ss'}.
\end{equation}
Here, we introduced a $\U(1)_\cN$ phase $\theta(x)$,
a three-component real scalar $d_\alpha(x)$ with $d^2_\alpha(x) = 1$,
an $\SO(3)_\cR$ matrix $R_{ab}(x)$,
and $\phi_0 = (1/\sqrt{2},i/\sqrt{2},0)^T$.
$d_\alpha(x)$ transforms in the adjoint representation under $\SU(2)_\cS$,
while $\theta(x)$ and $R(x)$ transform in the following ways under $e^{i \lambda(x)}\in \U(1)_\cN$ and $Q(x)\in\SO(3)_\cR$:
\begin{subequations}
    \begin{align}
        \theta(x) &\ra \theta(x) + \lambda(x),
        \\
        R(x) &\ra Q(x)R(x).
    \end{align}
\end{subequations}
Suppose that the fluctuations are turned off, i.e., $\theta(x) = \theta_0$, $R(x) = \1_3$, and $d_\alpha(x) = \delta_{\alpha 3}$.
Then, the order parameter is invariant under the spin rotation $e^{i \sigma_z \eta} \in \U(1)_\cS \subset \SU(2)_\cS$ as well as under the simultaneous rotations of $\U(1)_\cN$ by $\lambda/2$ and $\SO(3)_\cR$ by $- \lambda$ around $l_0 = (0,0,1)^T$. These are the unbroken symmetries $\U(1)_{\cN - \cR}$ and $\U(1)_{\cS}$ in Eq.~\eqref{eq: ssb}. 
Note that the order parameter is also invariant under $\theta(x) \ra \theta(x) + \pi/2$ and $d_\alpha(x) \ra -d_\alpha(x)$.
This discrete unbroken symmetry reduces the number of possible terms at the next-to-leading order.

While $d_\alpha(x)$ describes the two Nambu-Goldstone bosons in the spin sector,
$\theta(x)$ and $R(x)$ have $4$ components and are redundant corresponding to the unbroken symmetry $\U(1)_{\cN - \cR}$.
This redundancy in the $\U(1)_\cN \times \SO(3)_\cR$ sector is removed by the invariance of the order parameter~\eqref{eq: coset rep} under the local transformation:
\begin{subequations}
    \begin{align}
        \theta(x) &\ra \theta(x) + \lambda(x)/2,
        \\
        R(x) &\ra R(x)e^{-i L_3 \lambda(x)},
    \end{align}
\end{subequations}
with $(L_c)_{ab} = -i\epsilon^{abc}$ being a generator of $\SO(3)_\cR$.

The covariant derivatives for these dynamical fields take the forms of
\begin{subequations}
    \begin{align}
        D_\mu \theta(x) &= \partial_\mu \theta(x) - A_\mu(x),
        \\
        D_\mu R(x) &= (\partial_\mu - i w_\mu(x))R(x),
        \\
        D_\mu d_\alpha(x) &= \partial_\mu d_\alpha(x) + \epsilon^{\alpha\beta\gamma}\delta B^\beta_\mu(x) d_\gamma(x).
    \end{align}
\end{subequations}
Here, $w_\mu(x)$ is the spin connection, which is a gauge field for the $\SO(3)_\cR$ symmetry defined as~\cite{Hoyos:2013eha,Fujii:2016mbc}
\begin{subequations}
    \begin{align}
        w_t(x) &= \left[ 
            - \frac{1}{2}\epsilon^{abc}e^{aj}(x)\partial_te^b_j(x)
            +\frac{1}{2m}\epsilon^{abc}D_a A_b(x)
            \right]L_c, \label{eq: spin connection t}
        \\
        w_i(x) &=
            - \frac{1}{2}\epsilon^{abc}e^{aj}(x)
            (\partial_ie^b_j(x) - \Gamma^k_{ij}(x)e^b_k(x))
            L_c,
    \end{align}
\end{subequations}
where $\Gamma^k_{ij}(x)$ denotes the Christoffel symbol of the spatial manifold.
The magnetic field of $A_\mu(x)$ appears in Eq~\eqref{eq: spin connection t}, so that the spin connection transforms as a vector under the nonrelativistic diffeomorphism. This is the most crucial consequence of the nonrelativistic diffeomorphism invariance.

The $\SU(2)_\cS$ sector of the effective field theory can be constructed by combining $d_\alpha(x)$ and its covariant derivative $D_\mu d_\alpha(x)$.
On the other hand, we take the coset construction approach~\cite{Coleman:1969sm, Callan:1969sn} to build the $\U(1)_\cN \times \SO(3)_\cR$ sector.
The minimal combinations invariant under the internal symmetries are the Maurer-Cartan forms:
\begin{equation}
    D_\mu \theta(x),
    \ \ \ 
    X^a_\mu(x) = \frac i2\epsilon^{abc}[- i R^{-1}(x) D_\mu R(x)]_{bc}.
\end{equation}
We parametrize the fluctuations in the coset $[\U(1)_\cN \times \SO(3)_\cR]/\U(1)_{\cN - \cR}$ as
\begin{subequations}\label{eq: velocity}
    \begin{align}
        mv_\mu(x) &= D_\mu \theta(x) + \frac{l^a_0X^{a}_\mu(x)}{2},
        \label{eq: velocity}
        \\
        X^a_{\pp \mu}(x) &= X^a_\mu(x) - l^a_0 l^b_0 X^b_{\mu}(x).
    \end{align}
\end{subequations}
While the latter transforms as a vector under $\SO(2)_{\cN-\cR} \simeq \U(1)_{\cN - \cR}$, the former is invariant and called the velosity vector because it transforms under the nonrelativistic diffeomorphism as 
\begin{subequations}\label{eq: velocity transf}
    \begin{align}
        v_{t}(x) &\ra v_{t}(x)-\xi^k(x)\partial_kv_{t}(x)-\xi^k(x)\partial_tv_{k(x)},
        \\
        v_{i}(x) &\ra v_{i}(x)-\xi^k(x)\partial_kv_{i}(x)-\xi^k(x)\partial_iv_{k}(x)
        \nonumber\\
        &\quad + g_{ik}(x)\partial_t \xi^k(x).
    \end{align}
\end{subequations}
This expression~\eqref{eq: velocity} is a natural generalization of that for the chiral superfluid in two dimensions~\cite{Hoyos:2013eha},
and it plays an essential role in Sec~\ref{sec: leading}.
Note that the unbroken symmetry shifts $X_{\pr \mu}(x) = l^a_0 X^a_{\mu}(x)$, so that this quantity appears only in the covariant derivatives via the minimal coupling.

On the other hand, the Bogoliubov quasiparticles are gapless
at the points $k = \pm k_f l_0$ in the momentum space with $k_f$ being the Fermi momentum.
Thus, the microscopic $^3$He field is decomposed as follows at low-energy:
\begin{equation}
    \psi(x) \sim e^{ i k_f l_0 \cdot x}\psi_+(x)
    +e^{ -i k_f l_0 \cdot x}\psi_-(x),
\end{equation}
and $\psi_{\pm 0}(x) = e^{- i \theta(x)}\psi_{\pm}(x)$ are employed as building blocks in the fermion sector.
The covariant derivative for the fermion fields is then given by
\begin{equation}
    D_\mu\psi_{\pm 0}(x) = (\partial_\mu - iX_{\pr\mu}(x)/2 - i B_\mu(x)) \psi_{\pm 0}(x).
\end{equation}

\subsection{Power counting}
The effective Lagrangian is constructed on the building blocks explained above and is expanded systematically in terms of their derivatives.
Thus, we must clarify our power counting scheme for the derivative expansion.
We regard
\begin{equation}\label{eq: zeroth order}
    \partial_\mu \theta(x),
    \ \ 
    R(x),
    \ \ 
    d_\alpha(x),
    \ \
    A_\mu(x),
    \ \
    H_\alpha(x),
    \ \
    e^a_i(x)
\end{equation}
as $\mathcal{O}(\partial^0)$ and
\begin{equation}\label{eq: first order}
    \partial_\mu,
    \ \ 
    \delta B_\mu(x),
    \ \
    \psi_{\pm0}(x)
\end{equation}
as $\mathcal{O}(\partial^1)$ quantities.
It is possible to consider $\partial_\mu\theta(x)$ as $\mathcal{O}(\partial^0)$
because the $\U(1)_\cN$ symmetry prohibits $\theta(x)$ from appearing without derivatives (see Refs.~\cite{Son:2005rv,Hoyos:2013eha,Moroz:2014ska,Fujii:2016mbc} for similar discussions).
$\psi_{\pm0}(x)$ are regarded as $\mathcal{O}(\partial^1)$ because the density of Bogoliubov quasiparticles must be small at zero temperature.
Consequently, fermion bilinear terms do not appear in the effective Lagrangian up to the next-to-leading order in the derivative expansion.

\subsection{Effective Lagrangian} \label{sec: leading}
Based on the power counting scheme explained above, we now construct the effective Lagrangian at $\mathcal{O}(\partial^0)$.
Only polynomials of 
\begin{subequations}
    \begin{align}
        \Theta(x) &= - m \left[ v_t(x) + \frac{v_a(x)v_a(x)}{2} \right],
        \\
        \Phi(x) &= \frac{1}{2}H^2_\alpha(x),
        \\
        \Psi(x) &= \frac{1}{2}[d_\alpha(x)H_\alpha(x)]^2
    \end{align}
\end{subequations}
are scalars under the nonrelativistic diffeomorphism. 
Thus, the most general effective Lagrangian is an arbitrary function of $\Theta(x)$, $\Phi(x)$, and $\Psi(x)$, which takes the form of
\begin{equation}\label{eq: leading order}
    \mathcal{L}^{(0)}(x) = f_0\left(\Theta(x),\Phi(x),\Psi(x)\right).
\end{equation}
This arbitrary function $f_0$ can be identified as the pressure of a ground state at the leading order~\cite{Son:2005rv}.
This is because we have $\Theta = \mu$ and a constant $d_{\alpha}$ for the ground state, so that the arbitraty function is related to the particle number density and the spin density as
\begin{subequations}
    \begin{align}
        \rho(\mu, H_\alpha) &= \frac{\partial f_0}{\partial \mu} (\mu, H_\alpha),
        \\
        s_\alpha(\mu, H_\alpha) &= \frac{\partial f_0}{\partial H_\alpha}(\mu, H_\alpha).
    \end{align}
\end{subequations}

On the other hand,
the following two terms at $\mathcal{O}(\partial^1)$ are allowed by the time-reversal and parity symmetries and the unbroken discrete symmetry:
\begin{subequations}
    \begin{align}
        \mathcal{L}^{(1)}_1(x) &= 
        f_1(\Theta(x),\Phi(x),\Psi(x))
        \nonumber \\
        &\quad \times \epsilon^{\alpha\beta\gamma}d_\alpha(x) H_\beta(x) (D_t + v^i(x) D_i)d_\gamma(x),
        \\
        \mathcal{L}^{(1)}_2(x) &= 
        f_2(\Theta(x),\Phi(x),\Psi(x)) (d_{\alpha}(x)H_\alpha(x))
        \nonumber \\
        &\quad \times i H_\alpha(x) (D_t + v^i(x)D_i) d_\alpha(x) ,
    \end{align}
\end{subequations}
where $f_1$ and $f_2$ are arbitrary functions.
The covariant derivative in the temporal direction must enter the effective Lagrangian in the combination $(D_t + v^i(x) D_i)$ because of the nonrelativistic diffeomorphism invariance.
Note that the derivatives of $H_\alpha(x)$ are neglected because we are eventually interested in situations with a constant magnetic field.
Finally, the effective Lagrangian up to the next-to-leading order in the derivative expansion reads
\begin{equation}
    \mathcal{L}_\mathrm{eff}(x)
    =   \mathcal{L}^{(0)}(x) + \sum^2_{n=1}\mathcal{L}^{(1)}_n(x).
\end{equation}

%---------------------------------------------
\section{Hall viscosity} \label{sec: hall}
\subsection{Universal expression}
To derive the Hall viscosity from the effective Lagrangian constructed above, we need to identify the orbital angular momentum density, which appears in the particle number current, and compute a part of the stress tensor in the A-phase.
The particle number current is obtained by differentiating the effective action
\begin{equation}\label{eq: effective action}
    S_\mathrm{eff} = \int d^4x \sqrt{g(x)} \mathcal{L}_\mathrm{eff}(x)
\end{equation}
with respect to the background $\U(1)_\cN$ gauge field $A_\mu(x)$.
When it is varied as $A_\mu(x) \ra A_\mu(x)+\delta A_\mu(x)$, we find 
\begin{equation}
    \delta S_\mathrm{eff}
    = \int d^4x \sqrt{g(x)} \left[ \rho(x)\delta A_t(x)+ J^i(x)\delta A_i(x) \right],
\end{equation}
where the particle number density and current are defined as
\begin{subequations}
    \begin{align}
        \rho(x) &= \frac{\partial f_0}{\partial \Theta}(\Theta(x),\Phi(x),\Psi(x))
        +\sum^2_{n=1} \rho_n(x)
        ,
        \\
        J^i(x) &= \rho(x) v^i(x) 
        + \sum^2_{n=1} J^i_n(x)
        + J^i_\mathrm{rot}(x)
        \\
        & = \rho(x) v^i(x) 
        + \sum^2_{n=1} J^i_n(x)
        + \frac{\varepsilon^{ijk}(x)D_j [\rho(x)l_k(x)]}{4m}.
        \label{eq: current}
    \end{align}
\end{subequations}
Here, the contributions from $\mathcal{L}^{(1)}_n(x)$ are denoted as $\rho_n(x)$ and $J_n(x)$ for simplicity and $\varepsilon^{ijk}(x) = \epsilon^{abc}e^{ai}(x)e^{bj}(x)e^{ck}(x)$ is defined.
We also introduce the $l$-vector $l_a(x) = R_{ab}(x) l^b_0$ and $\overline{l}_i(x) = \rho(x) l_i(x)/2$.
$\overline{l}_i(x)$ represents the orbital angular momentum density originating from the spinning Cooper pairs because it contributes to the orbital angular momentum as
\begin{equation}
        \overline{L}_i
        = m\int d^3x\, \varepsilon_{ijk}(x)x^j J^k_{\mathrm{rot}}(x)
        =\int d^3x\, \overline{l}_{i}(x).
\end{equation}

We then consider a variation with respect to the vierbein $e^a_i(x)$ to obtain the stress tensor,
which is defined as
\begin{equation}
    T^{ij}(x) = \frac{1}{\sqrt{g(x)}}\left[ 
        e^{aj}(x)\frac{\delta S_\mathrm{eff}}{\delta e^a_i(x)}
        +e^{bi}(x)\frac{\delta S_\mathrm{eff}}{\delta e^a_j(x)}
    \right].
\end{equation}
For the effective action~\eqref{eq: effective action},
the stress tensor takes the form of
\begin{equation}
    \begin{split}
        T^{ij}(x) &= g^{ij}(x) \mathcal{L}_\mathrm{eff}(x) - m \rho(x)v^i(x) v^j(x)
        \\
        &\quad +v^i(x) J^j(x) +v^j(x) J^i(x)
        \\
        &\quad - \eta^{ij;kl}_\mathrm{Hall}(x)V_{kl}(x)
        +\mathcal{O}(\partial^2).
    \end{split}
\end{equation}
Here, $V_{ij}(x) = (D_i v_j(x) + D_j v_i(x) + \partial_t g_{ij}(x))/2$ denotes the strain-rate tensor and the Hall viscosity $\eta^{ij;kl}_\mathrm{Hall}(x)$ is found to be
\begin{equation}\label{eq: hall viscosity}
    \begin{split}
    \eta^{ij;kl}_\mathrm{Hall}(x) = - \frac{\overline{l}_n(x)}{4}\Big[
        \varepsilon^{nil}(x)g^{jk}(x) + \varepsilon^{nik}(x)g^{jl}(x)
        \\ {} +
        \varepsilon^{njl}(x)g^{ik}(x) + \varepsilon^{njk}(x)g^{il}(x)
        \Big].    
    \end{split}
\end{equation}
Thanks to the antisymmetric property $\eta^{kl;ij}_\mathrm{Hall}(x) = -\eta^{ij;kl}_\mathrm{Hall}(x)$, the Hall viscosity is dissipationless.
It should be remarked that the leading contribution to the Hall viscosity is completely fixed by the intrinsic orbital angular momentum density $\overline{l}_i(x)$.
The expression~\eqref{eq: hall viscosity} takes the same form as the B-phase~\cite{Fujii:2016mbc}, and it is universal because our derivation depends only on the symmetries.

\subsection{Experimental observable}
Finally, let us clarify the implication of the Hall viscosity.
We turn off the background fields and prepare the situation, 
where the particle number density is constant (i.e., $\rho(x) = \rho_0$) and the $l$-vector points at the $x^3$ direction (i.e., $l(x) = (0,0,1)$).
We then perturb the uniform ground state as $\rho(x) = \rho_0 + \delta \rho(x)$ and $J_i(x) = \delta J_i (x)$
and linearize the particle number and momentum conservation equations to obtain
\begin{subequations}\label{eq: eom}
    \begin{align}
        \partial_t \delta \rho(x) + \partial_i \delta J^i(x) &= 0,
        \\
        m \partial_t \delta J^i (x) + m c^2 \partial_i \delta \rho(x)
        &= \eta^{ij;kl}_\mathrm{Hall}\partial_j \partial_k \delta J_l(x)/\rho_0.
    \end{align}
\end{subequations}
Here, the speed of sound $c$ is defined as $m c^2 = \partial P/\partial \rho$.
In particular, for the density modulation $\delta \rho(x) = \delta\tilde{\rho}\,e^{-i \omega t + ik x^1}$, the set of equations~\eqref{eq: eom} yields the sound-wave solution:
\begin{subequations}
    \begin{align}
        \delta J_1(x) &= \frac{ \omega }{k} \delta\tilde{\rho}\,e^{-i  \omega t + ik x^1},
        \\
        \delta J_2(x) &= -\frac{i}{4m} k \delta \tilde{\rho}\,e^{-i \omega t + ik x^1},
    \end{align}
\end{subequations}
with the dispersion relation $ \omega^2=c^2k^2$.
In contrast to the sound wave in $s$-wave superfluids, the sound wave in the A-phase has not only the longitudinal component but also the transverse one because of the Hall viscosity~\eqref{eq: hall viscosity}.
Namely, the sound wave is elliptically polarized.
The ratio between the longitudinal and transverse components is proportional to the Hall viscosity and estimated as 
\begin{equation}
    \frac{|\delta J_2|}{|\delta J_1|} = \frac{\hbar \omega}{4 m c^2} \sim 6 \times 10^{-8},
\end{equation}
for representative values of $\omega \sim 1\ \mathrm{MHz}$ and $c \sim 300\ \mathrm{m/s}$~\cite{Wheatley:1975tf} as well as $\hbar = 1.05 \times 10^{-34}\ \mathrm{J \cdot s}$ and $m = 5.01 \times 10^{-27}\ \mathrm{kg}$.
This ratio is found to be five orders of magnitude larger than that in the B-phase under a reasonable magnetic field~\cite{Fujii:2016mbc}, so that its experimental observation is more feasible in the A-phase.

%---------------------------------------------
\section{Summary} \label{sec summary}
In this paper, we constructed the effective field theory describing low-energy dynamics in the A-phase of superfluid $^3$He at zero temperature under a sufficiently large magnetic field.
This approach is based solely on the symmetries and the systematic derivative expansion, so that it is nonperturbative, model-independent, and qualitatively reliable.
To this end, we turned on the background gauge fields and spatial metric so as to expose the nonrelativistic diffeomorphism invariance, which is regarded as a local extension of the Galilean invariance.
We then established the most general effective Lagrangian up to the next-to-leading order in the derivative expansion respecting the nonrelativistic diffeomorphism invariance as well as the $\U(1)_\cN \times \SO(3)_\cR \times \SU(2)_\cS$ symmetries.
In particular, the nonrelativistic diffeomorphism invariance uniquely fixes the temporal component of the spin connection, which leads to the universal expression for the Hall viscosity.
We found it to be proportional to the orbital angular momentum density and five orders of magnitude larger than that in the B-phase under a reasonable magnetic field.
Therefore, the experimental observation of the Hall viscosity is more feasible in the A-phase, for example, by measuring the induced elliptic polarization of sound waves.
It is indeed worthwhile if our prediction could be confirmed in the laboratory.

\acknowledgments
This work was supported by JSPS KAKENHI Grants No.\ JP20J13415, JP19J13698, and JP18H05405.

\bibliography{3He-A-phase}
\end{document}